# Heter-LP: A heterogeneous label propagation algorithm and its application in drug repositioning

Maryam Lotfi Shahreza
*Department of Electrical and Computer Engineering,
Isfahan University of Technology,
Isfahan 84156-83111, Iran*
`maryam.lotfi@ec.iut.ac.ir`

Nasser Ghadiri *
*Department of Electrical and Computer Engineering,
Isfahan University of Technology,
Isfahan 84156-83111, Iran*
`nghadiri@cc.iut.ac.ir`

Seyed Rasul Mossavi
*Department of Computer Engineering,
Amirkabir University of Technology
Tehran 15916-34311, Iran*
`srm@aut.ac.ir`

Jaleh Varshosaz
*Department of Pharmaceutics, School of Pharmacy and Pharmaceutical Science,
Isfahan University of Medical Sciences, Isfahan, Iran*
`varshosaz@pharm.mui.ac.ir`

James Green
*Department of Systems and Computer Engineering
Carleton University
Ottawa, Canada*
`jrgreen@sce.carleton.ca`

*Abstract*— Drug repositioning offers an effective solution to drug discovery, saving both time and resources by finding new indications for existing drugs. Typically, a drug takes effect via its protein targets in the cell. As a result, it is necessary for drug development studies to conduct an investigation into the interrelationships of drugs, protein targets, and diseases. Although previous studies have made a strong case for the effectiveness of integrative network-based methods for predicting these interrelationships, little progress has been achieved in this regard within drug repositioning research. Moreover, the interactions of new drugs and targets (lacking any known targets and drugs, respectively) cannot be accurately predicted by most established methods.

*(\*) Corresponding author. Address: Department of Electrical and Computer Engineering, Isfahan University of Technology, Isfahan, Iran. Phone : +98-31-3391-9058, Fax: +98-31-3391-2450, Alternate email: nghadiri@gmail.com*

In this paper, we propose a novel semi-supervised heterogeneous label propagation algorithm named Heter-LP, which applies both local as well as global network features for data integration. To predict drug-target, disease-target, and drug-disease associations, we use information about drugs, diseases, and targets as collected from multiple sources at different levels. Our algorithm integrates these various types of data into a heterogeneous network and implements a label propagation algorithm to find new interactions. Statistical analyses of 10-fold cross-validation results and experimental analysis support the effectiveness of the proposed algorithm.

# 1   Introduction

The process of finding additional indications for existing drugs is known as Drug Repositioning (DR). DR is a way to save time and costs when compared to the *de novo* drug development process. Computational methods can guide wet lab experimental design by narrowing the scope of candidate targets to accelerate drug discovery and can provide supporting evidence for experimental results.

Recent research shows that simultaneous use of the three concepts of diseases, drugs, and targets together leads to better results for drug repositioning [1]. Specifically, the application of network-based approaches in the fields of genomics, transcriptomics, proteomics, and systems biology have potential to improve drug development. These improvements will help decrease the time between lead development and drug marketability [2].

Advances in the biological sciences over the past several decades have resulted in the generation of large amounts of molecular data at the level of the genome, transcriptome, proteome, and metabolome, motivating the application of data-driven approaches. One of the most frequently used data-driven approaches is network modeling. However, each type of molecular data presents a limited view of the biological system. Hence, the combination of multiple data types can provide a more enriched and complete picture. Multilevel systems integration has been successfully applied to combine these heterogeneous data types into integrated networks in other application areas; for example, such integration has enabled efficient modeling of metabolic perturbation [3], [4], [5]. The use of heterogeneous networks in DR is motivated by the fact that drugs tend to take effect via interaction with one or more protein targets within a cell. Therefore, it is necessary to consider drugs, protein targets, and diseases simultaneously to investigate their inter-relationships.

The research objective of this paper is to provide a computational framework to facilitate drug repositioning tasks based on a heterogeneous network, generated by integrating drug, disease, and target information in different levels.

Specifically, this work includes two components. (1) *Construction of a heterogeneous network*: This network is composed of six networks: the drug similarity network, disease similarity network, target similarity network, known drug-disease associations, known drug–target interactions and known disease-target interactions. (2) *Algorithmic prediction of different potential interactions*: We here develop a heterogeneous label propagation algorithm to predict potential drug–target interactions, drug-disease associations, and disease-target interactions by integrating multi-source information. The reason for this choice is that heterogeneous network label propagation is an effective and efficient technique to utilize both local and global features in a network for semi-supervised learning [6]. We introduce a novel semi-supervised heterogeneous label propagation algorithm named Heter-LP. The Heter-LP algorithm consists of the following steps: (1) Data collection and preparation, (2) construction of similarity matrices and association matrices, and (3) label propagation. Then we use Heter-LP to develop a new drug repositioning method, by performing label propagation to integrate different levels of biological information and apply an optimization algorithm to find new drug-target interactions. The evaluation is based on a 10-fold cross-validation experiment design and we analyze the results using performance metrics such as the Area Under the Receiver Operating Characteristic Curve (AUC) and Area Under the Precision-Recall curve (AUPR).

The goals of this research are to identify putative candidates for drug repositioning, to further improve the prediction accuracy of drug-target interactions, and to discover useful drug-disease and disease-target associations. In this application, the inputs to our algorithm are three similarity matrices and three association matrices which are generated using three different levels of information (molecular originated profiles, molecular activity information, and phenotypic properties). The primary output will be three matrices representing drug-target interactions, drug-disease associations, and disease-target associations. The final output is a ranked list of candidates for drug repositioning. Secondary outputs include new similarity matrices for drugs, diseases, and targets which could have application in clustering for example.

Unlike existing methods, our proposed method can predict interactions of *new* drugs (where a drug has no known target) and *new* targets (where a target has no known drug). Moreover, in our approach, there is no requirement for negative training exemplars. The other benefit of the proposed method is its ability to predict both trivial and non-trivial interactions. We believe that meaningful and efficient integration of information is achieved due to our use of an appropriate structural network model and suitable label propagation algorithm. Moreover, the pre-phase projection phase enriches the algorithm. The statistical and experimental analysis will demonstrate these claims.

The paper is organized as follows. Section 2 presents the related work. A complete description of our proposed method is provided in Section 3. Section 4 proves the convergence of the method. In Section 5, the regularization framework is presented. The performance evaluation and a brief description of computational and time complexity of the algorithm is presented in Section 6. Finally, Section 7 provides a summary of the research.

## 2  Related Work

Gene expression patterns change systematically in response to disease processes. Transcriptome data provide a snapshot of such whole-genome dynamics and can provide insights into the mechanism of action of drugs [7]. Differential gene expression analysis is an effective way to identify genes that lead to disease. In this regard, some drug repositioning methods have been developed based on gene expression analysis, such as [8], [9], [10], [11], [12], [13], [14]. In spite of observed good performance, there are limitations associated with these methods that compare gene expression signatures. First, the set of drugs and diseases included in current databases of gene expressions are limited, so these methods are restricted to the subset of known diseases (such as particular kinds of cancers). So it seems that we cannot rely solely on such limited data and we need more data sources to complement them. Second, results coming from cell lines cannot always be extrapolated to *in vivo* tissues.

Protein-protein interaction (PPI) is often used as the basis for drug target identification since the PPI network provides the context in which the target protein operates. Here, there is an underlying assumption that the proteins targeted by similar drugs tend to be functionally associated and be close in the PPI network. Here, a *similar* drug refers to a drug that has "similar therapeutic effects" [7]. Some of the most important drug repositioning methods based on PPI include refs [15] and [16]. Despite demonstrated success in repositioning drugs using PPI networks, there are also some limitations. First, the required PPI data are noisy and incomplete, and the extracted networks are incomplete and biased [17]. Second, like gene regulatory networks, there is no simple mapping between a simulated network and a living organism's actual response [17].

By using a metabolic network, several important physiological properties of a cell could be extrapolated. So metabolic networks can also be used to predict drug targets. Two drug repositioning methods based on metabolic networks are described in [18] and [19].

The identification of novel drug-target interactions (DTI) is the basis for drug discovery and design, and accurate prediction of drug targets is a key to effective drug repositioning. Many drugs are non-specific and show reactivity to additional targets besides the primary targets. Although these off-target effects often leading to unwanted side effects (discussed below), these one-drug-multiple-target data can also be leveraged by DR methods. Motivated by this, researchers have developed many methods based on drug-target interactions, including [20], [21], [22], [23], [24], [25], [26], [27], [28], [29], [30], [31], [32], [33], [34]. Some of these methods are based strictly on DTI, while others are extended to leverage additional data such as protein-protein similarity and drug-drug similarity[1], often leading to increased performance. For example, when a given drug has several known targets, additional candidate targets can be ranked by calculating the similarity between candidate targets and known targets. However, in the case where a drug has no known target, (i.e. a new drug), computing target similarity is not possible. Hence, drug similarity must be used instead. In this case, potential targets of this new drug are selected based on target information for similar drugs for which target data is available. Semi-supervised learning methods can address the problem of predicting interactions for new drugs (or new targets).

In addition to the previously mentioned methods, some drug repositioning methods leveraging drug-drug similarity include refs [35], [36]. One of the main limitations of drug repositioning based on chemical structure similarity of drugs is that many structures and chemical properties of known drug compounds are inaccurate. Furthermore, many physiological effects of a drug cannot be predicted by structural properties alone [37].

Computational assessment of similarities in molecular profiles is another approach for relating drugs to disease states for the purpose of repositioning [37], [38]. The role of molecular profile could be described as a signature of molecular activity after exposing a drug in a biological system. It may contain different measures such as a change in transcriptional activity. The similarity of these profiles could be used to establish useful relationships between drugs and diseases. There have been important methods of this type in the literature [39], [40], [41].

Many drugs induce some unintended effects in the living organism besides the primary desired effects, which constitute a drug's overall effect profile. Those wanted or unwanted behavioral or physiological changes in response to drug treatment can be measured as a drug's indication and side effects, respectively [17], [42]. We know that side effects are generated when a drug binds to off-targets, which perturb unexpected metabolic or signaling pathways. These off-targets may, in fact, lead to the identification of novel therapeutic targets [7], [43]. So pharmacological information associated with drugs provides an alternative way to predict drug targets; this approach has proven to be complementary with the commonly used molecular information. Ye et al. [43] tried to reposition drugs by statistical analysis of drug side effects. Zhang et al. [44] predicted adverse drug-drug interactions by integrating side effects with molecular information. The PREDICT [40] algorithm also used drug side effects to rank drug-disease associations.

---

[1] There are manymetrics to measure the similarity between two drugs. For example, some are based on similarity of biological effects and some are based on similarity of chemical structures of drugs and so on. In this paper, special kinds of these similarities will be clarify explicitly whenever needed, and in general word "similarity of drug" could be referred to anysimilarity.

There is some limitation for using of side-effect. First, the scarcity of drug adverse reaction information limits the application of this kind of approaches. In fact, side-effect-similarity approaches need a well-defined side effect profile for a drug while current disease and drug phenotype data are noisy and far from complete. Second, there is no side effect profile available for newly approved drugs [17], [7], [37]. Third, all drugs with similar effects do not possess similar targets necessarily. Fourth, there is no simple mapping between phenotype and mechanism of action. The living organism's genetic map, medication history, and other traits could effect on phenotypic outcomes of a drug. So, we could not conclude that a similar phenotype corresponds to the same mode of action [17], [7]. Finally, it should be noticed that side effect information could be confused by a patient's medication history, genotype, and other hidden factors [17].

In drug repositioning, it is assumed that if molecular pathophysiology of therapeutic effects of two drugs has sufficient commonalities, they are interchangeable. [37]. So to reposition drugs, we require computational strategies for finding molecular relationships between distinct disease pathologies. This approach has been leveraged for DR in [42].

Previous studies indicate that integrative analysis, where multiple lines of evidence are considered simultaneously, is a practical approach to finding the most probable candidates for drug repositioning. Recently, some novel integrative methods have been proposed. Wang et al. [45] integrated three data sources from the structure, activity, and phenotype levels using a kernel function. Their PreDr method then uses an SVM-based predictor to uncover unknown interactions between drugs and diseases. However, the lack of high confidence known negative instances restricted its performance. They chose the known drug-disease pairs as the positives and randomly selected a set of training negatives from the unlabelled data. Many of these unlabeled data are, in fact, undiscovered drug-disease relationships; this leads to incorrectly labeled negative data and an important limitation that also applies to the other integrative approaches described below.

Yamanishi et al. [46] developed a method to predict unknown drug–target interactions from chemical, genomic, and pharmacological data. This statistical method is based on supervised bipartite graph inference. Additionally, they investigated the topology of drug–target interactions networks.

Wang et al. [47] proposed an integrative framework based on the information flow-based method, named TL_HGBI (Triple Layer Heterogeneous Graph Based Inference). The TL_HGBI tried to calculate the weight[2] of disease-drug relations by drug-target relationships using an iterative algorithm to combine heterogeneous information from different sources.

In [1], Zhang et al. proposed a framework named SLAMS (Similarity-based LArge-margin learning of Multiple Sources). SLAMS integrates different drug information, such as chemical properties (compound fingerprints), biological properties (protein targets), and phenotypic properties (side-effect profiles) to predict drug-target interaction.

NRWRH [48] integrates three different networks (protein similarity network, drug similarity network, and the known drug–target interaction network) into a heterogeneous network. It then uses a random walk method for the prediction of drug–target associations. This methodology demonstrated good performance in predicting new interactions. However, random walking may lead to a locally optimal solution due to the sparseness of the drug–target interaction network.

While integrative methods have been shown to outperform other approaches, these methods still face some common limitations outlined below:

1) Most existing similarity-based prediction algorithms use only immediate similarities and don't consider transitivity of similarity. To address this deficiency, Zhang et al. [44] proposed a label propagation approach that considers higher-order similarity which could be useful in drug repositioning research and help us in this regard.
2) As mentioned above, in many methods negative drug-target interactions are selected randomly without experimental confirmation [33].
3) Interactions with new drugs (drugs without any known target) and new targets (targets without any known drug) cannot be predicted by some methods [49]. Semi-supervised learning methods could be useful in addressing this problem.
4) Most existing methods are based on one or two kinds of data (like chemical structure similarity of drugs or sequence similarity of protein targets). On the other hand, existing training samples of established methods are very few when compared with all available unlabeled data.

Our proposed method is a semi-supervised method without requiring negative training samples, and capable of utilizing information from unlabelled samples. Furthermore, it is applicable in the case of the new entity problem. Lastly, integration of different data types is used to improve the prediction accuracy.

## 3   Method

We here design a novel algorithm to predict drug repositioning by associating known drugs with new diseases, different diseases with new targets, and drugs with new targets.

---

[2] confidence of the existence of the relationship

This section will introduce the Heter-LP method, using drug repositioning as the illustrative example application. Subsection 3.1 presents the formal notations and settings used in the problem. Subsection 3.2 covers data collection. Subsection 3.3 is about the projection step of the algorithm. Subsection 3.4 explains the label propagation algorithm. Finally, in subsection 3.5, pseudo-code of the algorithm is presented.

The proposed data model consists of six parts, three homogeneous (1, 2, 3) and three heterogeneous (4, 5, 6) sub-networks: (1) Drug similarity network, (2) Disease similarity network, (3) Target similarity network, (4) Known drug-disease associations, (5) Known drug–target interactions, (6) Known disease-target associations.

A comprehensive description of using data for each part is represented in subsection 3.2. Our aim is to optimally integrate these different data sources and provide a ranked list of putative novel associations between drugs, diseases, and targets. Figure 1 is a schematic view of our heterogeneous network model and used datasets.

## 3.1 Notations and problem settings

Here we have three types of nodes: drugs, diseases, and targets. We have six different kinds of edges, each representing one type of similarity or association: drug similarity, disease similarity, target similarity network, known drug-disease associations, known drug–target interactions and known disease-target interactions.

Therefore, we have a heterogeneous graph $G = (V, E)$ with three homo-sub-networks and six hetero-sub-networks (Figure 1). The homo-sub-networks are defined as $G_i = (V_i, E_i)$ where $i$ is $1, 2, 3$ for drugs, diseases, and targets, respectively. The hetero-sub-networks are: $G_{i,j} = (V_i \cup V_j, E_{i,j})$ for $i,j = 1, 2, 3$.

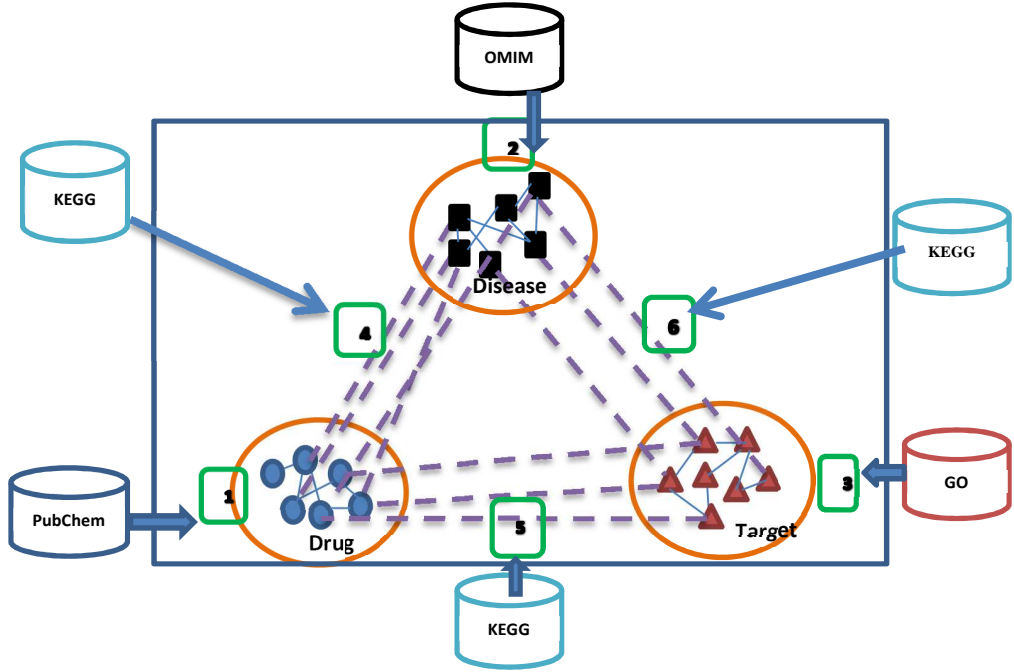

**Figure 1. Heterogeneous network model and used datasets**

Each $E_i$ is the set of edges between vertices in the vertex set $V_i$ of homo-sub-network $G_i$ and each $E_{i,j} \in V_i \times V_j$ is the set of edges connecting vertices in $V_i$ and $V_j$.

So in $G$, $V = \{V_1, V_2, V_3\}$ and $E = \{E_1, E_2, E_3\} \cup \{E_{1,2}, E_{1,3}, E_{2,3}\}$.

We represent the inputs of homo-subnetworks by one $n_i \times n_i$ affinity matrix[3] $A_i$, where $n_i$ is the number of vertices in corresponding homo-subnetwork and $A_i(k,k') \geq 0$ is the similarity between entity $k$ and $k'$. For example, the input drug network is represented by an $|V_1| \times |V_1|$ element square symmetric affinity matrix $A_1$, where $A_1(k,k') \geq 0$ is the similarity between drug $k$ and $k'$. For each hetero-subnetwork, there is a relation matrix $A_{i,j}$ with $|V_i|$ rows and $|V_j|$ columns. Each entry $A_{i,j}(k,k') \in \{0,1\}$ reflects the absence or existence of a relation between entity k and entity k', respectively. For example, the input drug-target network is represented by a $|V_1| \times |V_3|$, binary matrix, named $A_{1,3}$. A value of $A_{1,3}(k,k')=1$ indicates that there is a relation between drug $k$ and target $k'$, whereas a value of $A_{1,3}(k,k')=0$ indicates there is no relationship between drug $k$ and target $k'$. All $A_i$ and $A_{i,j}$ matrices must be normalized (once at initialization) to ensure convergence of the updates, so that each row sum is one. We used the "LICORS" package [50] in R for this task and denoted the normalized matrices as $S_i$ and $S_{i,j}$.

In Figure 2, we try to clarify the process using the workflow. A brief description of each part and the data preparation methods are represented below.

---

[3] All homogenous sub-networks are affinities symmetric.

## 3.2 Data preparation

In this section, we will study the characteristics of the datasets that are used in this research. Here we have three major concepts: disease, drug, and target. An intra-similarity matrix represents each concept, and three interaction matrices are describing the relation between each pair of these concepts.

We used a gold standard dataset provided by Yamanishi (2008) [24] to provide the opportunity to compare prediction accuracy with previous methods more accurately. Additionally, we also gathered several independent datasets to provide a more realistic experimental analysis. Further details are provided below.

### 3.2.1 Gold standard dataset

Yamanishi et al [24] provide a dataset contains drug, protein targets, and their interactions which are categorized by four groups of protein targets (Enzyme, GPCR, Ion Channel, and Nuclear Receptor). The primary resources of these data are KEGG[4], BRITE[5], BRENDA[6], SuperTarget[7], and DrugBank[8]. The similarity-by structure of drugs is computed by SIMCOMP on chemical substructures, and similarity of targets is calculated by a normalized version of Smith-Waterman score. Some of the most important characteristics of these data are represented in Table 1 of supplementary materials.

---

[4] http://www.kegg.jp
[5] http://www.genome.jp/kegg/brite.html
[6] http://www.brenda-enzymes.org/
[7] http://insilico.charite.de/supertarget/
[8] http://www.drugbank.ca

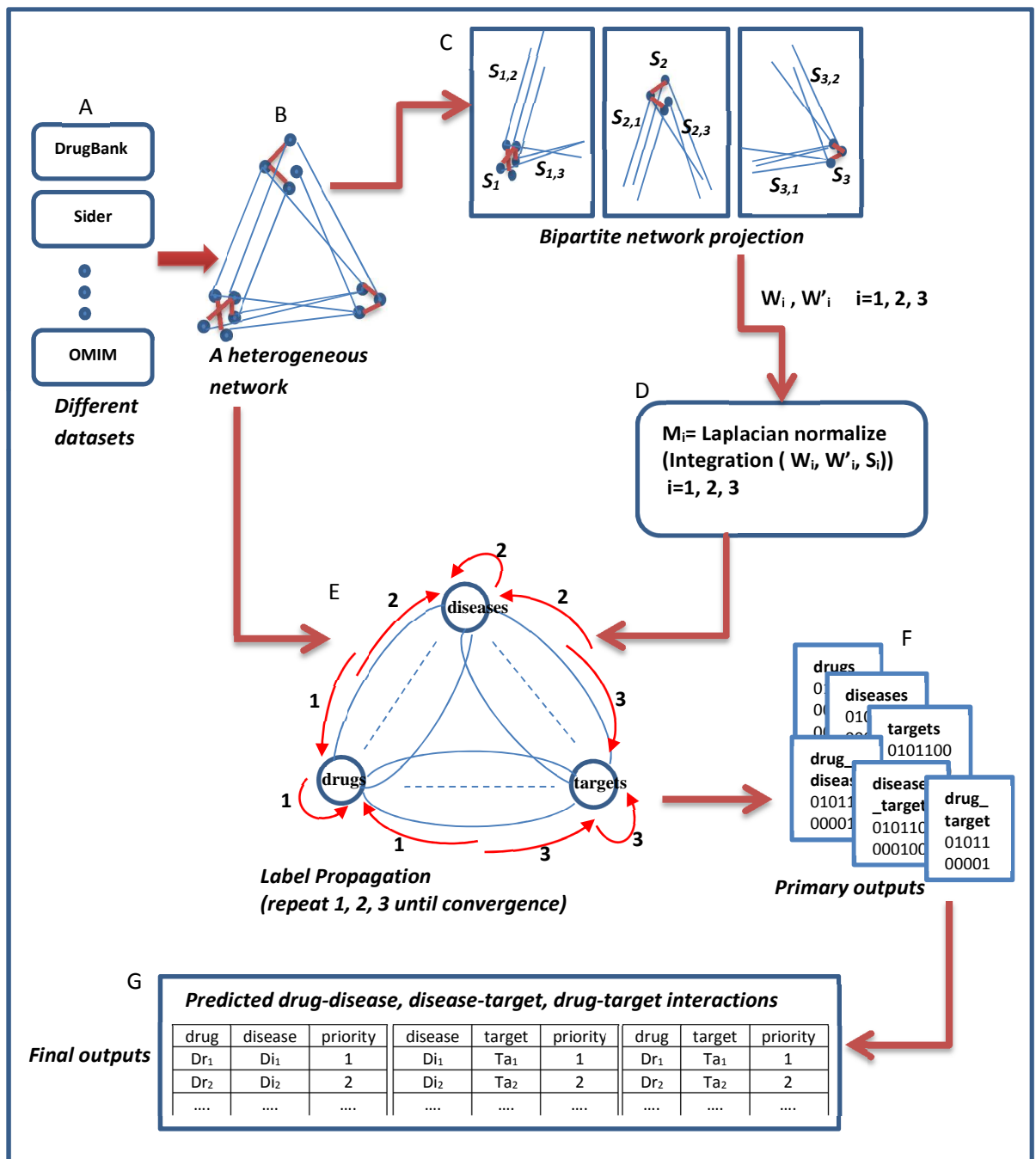

**Figure 2** The overall process workflow for Heter-LP algorithm. The alphabetic order could be used to find an appropriate understanding of the proposed method; **(A)** A schematic view of the datasets, explained in section 3.2 **(B)** The constructed network explained in section 3.1 and Fig. 1; **(C)** Our projection phase (equivalent to lines 2-7 in Algorithm section of represented pseudo code) explained in section 3.3.; **(D)** Equivalent to lines 8-10 in Algorithm section of represented pseudo code. In this part, we try to provide a laplacian normalization view of projection output.
**(E)** The label propagation phase of the algorithm (equivalent to lines 11-13 in Algorithm section of represented pseudo code) explained in section 3.4. Arrows with label 1 are equivalent to line 11 and so on. **(F)** Primary outputs are six matrices. Three of them are related to homo-sub-networks and three are related to hetro-sub-networks. Explanation is available in the end of section 3.5. **(G)** Final outputs are three sorted list of predicted interactions. Explanation is available in the end of section 3.5.

Though this dataset provides a useful benchmark for comparison of different methods, it has some limitations. In addition to being somewhat outdated, there is no information about associated diseases of drugs and targets in this dataset. Since the the concept of disease is a fundamental element in our model, we had to incorporate some extra data. For this purpose, we provided a list of related disease for each mentioned groups of gold standard datasets based on data provided by Li *et al.* [51] and DisGeNet [52], for drugs and targets respectively, and provide disease similarity from OMIM [53]. For example, we extracted all diseases related to drugs of Enzymes files of [24] from drug-disease lists of [51] in the form of a separate matrix and so on for other groups. Code written for this purpose is available upon request.

### 3.2.2 Independent test datasets

In addition to the gold standard dataset, we also build an updated heterogeneous dataset gathered from a number of datasets representing the three key concepts used by the model (drug, disease, and target) and their inter-relations. A brief description of each dataset is provided below.

**Homogeneous sub-graphs**

**G1.** *Chemical substructures:* It is believed that drugs with similar chemical structure carry out similar therapeutic function, thus, are likely treat common diseases [45].

In this experiment, the pairwise similarity of two drugs is calculated based on the *2D* chemical fingerprint descriptor of each chemical structure in PubChem. That is, each drug *d* is represented by a binary fingerprint *h(d)* in which each bit represents a predefined chemical structure fragment. Each *h(d)* is an 881-dimensional chemical substructure vector defined in PubChem. These data were taken from [54][9], which contains 888 drugs and 881 substructures, and the description of the 881 chemical substructures is available at PubChem's website. The pairwise chemical similarity between two drugs $d_1$ and $d_2$ is computed as the Tanimoto coefficient of their fingerprints using the "proxy" R package[10].

*Side effect similarities:* Many drugs have adverse effects in addition to their indication, referred to as side effects. It has been shown that more accurate drug-target prediction can be achieved by integrating these side effects with other information, such as chemical similarities [7]. SIDER [55] is the primary resource for side effect data. We found 888 drugs and 1385 side effects in this online dataset. These data also represented drug-side-effect relationships using a fingerprint matrix (like one described in chemical structure similarity). The similarity of side effects between two drugs $d_1$ and $d_2$ is computed as the Tanimoto coefficient of their fingerprints using the "proxy" R package. This distance is integrated with the chemical structure similarity matrix using the arithmetic mean to produce the $G_1$ sub-network. The combination of two drug-drug similarity matrices $g_1$, $g_2$ could be described by Eq. (1) below. Note that not all 888 drugs with structural fingerprint data had side effect information and vice-versa.

---

[9] http://cbio.ensmp.fr/yyamanishi/side-effect
[10] https://cran.r-project.org/web/packages/proxy/index.html

$$integration\ (g_1, g_2) = g$$

$$g(i,j) = \begin{cases} (g_1(i,j) + g_2(i,j))/2 & if\ entry(i,j)\ exist\ in\ g_1\ and\ g_2 \\ 2g_1(i,j)/3 & if\ entry(i,j)\ not\ exist\ in\ g_2 \\ 2g_2(i,j)/3 & if\ entry(i,j)\ not\ exist\ in\ g_1 \end{cases} \quad (1)$$

**G2.** *Semantic similarity of disease phenotypes:* Text mining techniques were utilized to classify human phenotypes contained in the Online Mendelian Inheritance in Man (OMIM) database [53]. The phenotype similarity data are accessible through the website at http://www.cmbi.ru.nl/MimMiner/. A matrix of pairwise semantic similarity between diseases is available from their website which was constructed such that each entry is calculated based on the number of co-occurring MeSH (Medical Subject Headings) terms in the specialist descriptions of each disease pair from the OMIM database. In other words, each row (or column) is the phenotype similarity profile for a single disease. The number of diseases in this matrix is 4784.

**G3.** *Gene Ontology-based Sematic Similarity Measures:* A drug target is a human protein, whose activity is modified by a drug, resulting in a desirable therapeutic effect [1]. Semantic similarity measures can be used to estimate the functional similarities among Gene Ontology terms and gene products. The pairwise semantic similarity protein pairs (drug targets) was computed using the "GOSemSim" package in R [56]. Here, we have 1537 protein targets.

**Heterogeneous sub-graphs**

Wu et al. [41] provided a useful list of relations between drugs, diseases, and protein targets extracted from KEGG. We divided these data into three separate lists of drug-disease, disease-target, and drug-target relationships and then, by using the "reshape2" [57] R package, converted them into corresponding fingerprint matrices. So we have a drug-disease matrix, $G_{1,2}$, containing 584 drugs and 203 diseases with 1041 relations between them; a drug-disease matrix, $G_{1,3}$, containing 3592 drugs and 1504 targets with 11610 relationships between them; and a disease-target matrix, $G_{2,3}$, containing 1087 diseases and 2255 targets with 3296 relationships. The number of each type of entity in different sub-networks is shown in Table 2 of supplementary materials.

In the present work, drugs and diseases are referred to by name, while targets are referred to by KEGG ID. Conversion between IDs is accomplished using online tools like Synergizer[11].

### 3.3 Projection

Bipartite networks have received considerable attention in the research community in different scientific areas. A bipartite network consists of two kinds of vertices (*X*, *Y*), and their edges are only allowed to connect to two vertices from different sets. Bipartite networks can model many different systems. To find a direct association between a particular set of vertices, we can compress them using "one-mode projection". One-mode projection onto *X* results in a network containing only *X* vertices. In this newly generated network, two vertices are connected if they have at least one common neighbor in the primitive network. This projection can be weighted or unweighted; the weighted type is usually preferred. We will use a weighted one-mode projection technique based on the method represented by Zhou et.al [58].

If *A* is an adjacency matrix between two vertex sets *X* and *Y*, we can project *A* onto *X* by Eq. (2):

$$W_{ij} = \frac{S_{ij}}{K(a_i)^{1-\lambda} K(a_j)^{\lambda}} \sum_{l=1}^{m} \frac{a_{il} a_{jl}}{K(a_l)} \quad (2)$$

Here *S* is similarity matrix of vertex set *X*; *K(x)* is equal to the degree of vertex *x* in matrix *A*; and *λ* is diffusion parameter of the projection. If *A* is an n-by-m matrix, and *S* is n-by-n, the resulting matrix *W* will be n-by-n as well.

As shown in **Error! Reference source not found.** and described in bipartite network projection section, we consider the primitive heterogeneous network as having three partitions. We apply our one-mode projection technique two times in each part separately. So we will have six different projection matrices. Each matrix can be used as a topological similarity matrix in the next section to improve the label propagation accuracy. We call them topological similarity matrices because of their ability to show different relations by weighted edges between one type vertices directly.

### 3.4 Label propagation

As noted in Section 1, we develop a heterogeneous label propagation algorithm to predict different types of the potential interactions in the network. In the naïve Label Propagation (LP) algorithm, there is an undirected weighted network with *n* nodes, of which are labeled, and the goal is to estimate the labels of the unlabeled nodes. Here, in each iteration, we have only one labeled node, and we attempt to predict the labels of others. In drug-drug homogeneous matrix the predicted labels would indicate similarities of drugs, and in drug-target heterogeneous matrix the predicted labels would indicate the existence of an interaction between corresponding drug and target.

---

[11] http://llama.mshri.on.ca/synergizer/translate/

The label propagation algorithm is closely related to the Random Walk (RW) algorithm. There are two major differences between LP and RW: (i) LP fixes the labeled points, and (ii) the solution of LP is an equilibrium state, while RW is dynamic. The LP algorithm is mathematically identical to Random Walk with Restart (RWR) algorithm if some constraints are applied to similarity matrix of RWR [32].

In this study, we require an algorithm to propagate label information across a heterogeneous network of drugs, targets, and diseases. A heterogeneous network is a network that consists of several sub-networks of different types of vertices and edges. For instance, in a drug-target hetero-network, there are two kinds of nodes and three kinds of edges. Edges between two drugs and between two targets are weighted as an explanation of their similarity. The edges between a drug and a target are un-weighted edges which are here binary, indicating the presence or absence of a relationship between the drug and the target. Most graph-based label propagation algorithms propagate label information only on a single network or homogenous network, which are not suitable for spreading label information across heterogeneous networks. In this regard, Hwang et al. [59] proposed a heterogeneous label propagation algorithm, named MINProp. This method sequentially propagates the label on each sub-network. Another heterogeneous label propagation algorithm is LPMIHN [32], with the primary purpose of inferring potential drug–target interactions using a heterogeneous network. The LPMIHN algorithm propagates labels on each homogenous sub-network separately, and then the interactions between the two homogenous sub-networks are used only as extra information in the form of a similarity matrix.

Here we apply label propagation on each sub-network using the existing information derived from the other sub-networks. The process repeats until convergence.

In brief, the inputs of the proposed algorithm include the similarity matrices and interaction matrices. In each iteration, we aim to find the relations between each pair of entities using these inputs. We initially set the label of a particular entity to one and all others to zero. This label information is propagated through the entire network to determine the relationship between the investigated entity and all others as newly assigned labels emerge. These new labels are saved in three vectors and, before the next iteration, are saved in specific matrices. Finally, we sort these output matrices from largest to smallest value of achieved label and determine the most important relations as the top scoring elements in each matrix.

### 3.5 Implementation

To clarify details of the proposed method, the pseudo code of our Heter-LP algorithm is presented below as Algorithm1.

**Algorithm 1 Heter-LP**

**Input**

1) $\sigma$: total convergence threshold
2) $\sigma'$: homogenous convergence threshold
3) $\alpha$: diffusion parameter of label propagation
4) $\lambda$: diffusion parameter of projection
5) $y_1, y_2, y_3$: vectors of initial label values
6) $S_1, S_2, S_3$: homo-subnetwork matrices
7) $S_{1,2}, S_{1,3}, S_{2,3}$: hetero-subnetwork matrices
8) drugs list ($n_1$ is the number of total drugs)
9) diseases list ($n_2$ is the number of total diseases)
10) targets list ($n_3$ is the number of total targets)

**Output**

1) $F_1, F_2, F_3$: homo-subnetwork matrices of final label values
2) $F_{1,2}, F_{1,3}, F_{2,3}$: hetero-subnetwork matrices of final label values

**Algorithm**

1. $F_k=0, F_{k,k'}=0$ for all $k,k'=1,2,3$

//Projection

2. $W1_{n1*n1}$ = projection of $S_{1,2}$ on $S_1$
3. $W'1_{n1*n1}$ = projection of $S_{1,3}$ on $S_1$
4. $W2_{n2*n2}$ = projection of $S_{1,2}$ on $S_2$
5. $W'2_{n2*n2}$ = projection of $S_{2,3}$ on $S_2$
6. $W3_{n3*n3}$ = projection of $S_{1,3}$ on $S_3$
7. $W'3_{n3*n3}$ = projection of $S_{2,3}$ on $S_3$

//Integration of similarity matrix with projected matrices

8. $M_1$=`NormalizeSumOf`($S_1, W_1, W'_1$)
9. $M_2$=`NormalizeSumOf`($S_2, W_2, W'_2$)
10. $M_3$=`NormalizeSumOf`($S_3, W_3, W'_3$)

// label propagation

11. **for** $i=1..\ y_1.length$

    11.1) $y_1[i]=1$, $y_1[j]=0$ **for all** $j \neq i$

    11.2) $y_2=y_3=0$

    11.3) $f_1=y_1, f_2=y_2, f_3=y_3$ // vectors of final label values

    11.4) `LabelPropagation`($f_1, f_2, f_3$)

    11.5) update $F_1$, $F_{1,2}$, $F_{1,3}$

 12.  **for** $i=1..\ y_2.length$

   *12.1)* $y_2[i]=1$, $y_2[j]=0$ **for all** $j \neq i$

   *12.2)* $y_1=y_3=0$

   *12.3)* $f_1=y_1, f_2=y_2, f_3=y_3$

   *12.4)* `LabelPropagation`$(f_1, f_2, f_3)$

   *12.5)* update $F_2$, $F_{2,1}$, $F_{2,3}$

 13.  **for** $i=1..\ y_3.length$

   *13.1)* $y_3[i]=1$, $y_3[j]=0$ **for all** $j \neq i$

   *13.2)* $y_1=y_2=0$

   *13.3)* $f_1=y_1, f_2=y_2, f_3=y_3$ // vectors of final label values

   *13.4)* `LabelPropagation`$(f_1, f_2, f_3)$

   *13.5)* update $F_3$, $F_{3,1}$, $F_{3,2}$

 14.  $F_{1,2}=$mean $(F_{1,2}$, transpose$(F_{2,1}))$

 15.  $F_{1,3}=$mean $(F_{1,3}$, transpose$(F_{3,1}))$

 16.  $F_{2,3}=$mean $(F_{2,3}$, transpose$(F_{3,2}))$

 17.  **return** $F_1, F_2, F_3, F_{1,2}, F_{1,3}, F_{2,3}$

---

`NormalizeSumOf(S,W,W')`

1. $d=0$ //a vector with S.numberOfRows length
2. **for** i=1..S.**numberOfRows**
   - 2.1. **for** j=1..S.**numberOfColumns**
     - 2.1.1. $M[i,j]=S[i,j]+W[i,j]+W'[i,j]$
     - 2.1.2. $d[i]=d[i]+M[i,j]$
   - 2.2 **if** $(d[i]==0)$ $d[i]=1$
3. **for** i=1..M.**numberOfRows**
   - 3.1. **for** j=1..M.**numberOfColumns**
     - 3.1.1. **if** $(i==j)$ $M[i,j]=1$
     - 3.1.2 **else if** ( i! = j **and** $M[i,j]! = 0$ ) $M[i,j]= \frac{M[i,j]}{\sqrt{d[i]d[j]}}$
     - 3.1.3 **else** $M[i,j]=0$
4. **return** (M)

---

`LabelPropagation`$(f_1, f_2, f_3)$

1. **repeat** (steps 2-11)

 //drug

2. $f_1 old = f_1$

3. $y'_1 = (1-\alpha_1)f_1 + \alpha_1(S_{1,2}*f_2 + S_{1,3}*f_3)$ //$f_1$ is equal to $y_1$

4. $f_1 = (1-\alpha_1)y'_1 + \alpha_1*M_1*f_1$

//disease

5. $f_2old = f_2$

6. $y'_2 = (1-\alpha_2)f_2 + \alpha_2((S_{1,2})^T*f_1 + S_{2,3}*f_3)$ //$f_2$ is equal to $y_2$

7. $f_2 = (1-\alpha_2)y'_2 + \alpha_2*M_2*f_2$

//target

8. $f_3old = f_3$

9. $y'_3 = (1-\alpha_3)f_3 + \alpha_3((S_{1,3})^T*f_1 + (S_{2,3})^T*f_2)$ //$f_3$ is equal to $y_3$

10. $f_3 = (1-\alpha_3)y'_3 + \alpha_3*M_3*f_3$

11. **while** ($||f_1-f_1old||>\sigma$ or $||f_2-f_2old||>\sigma$ or $||f_3-f_3old||>\sigma$ )

In Algorithm 1, first, we set all labels to zero. During the projection phase, we project interactions onto similarity matrices by equation (2). We will have six projected matrices which will be integrated with corresponding similarity matrices in lines 8-10. In label propagation phase we have three iterative loops (lines 11-13). In each loop, we set one of the original labels equal to one and all others to zero. The label propagation function is applied, and output matrices are updated. The primary output consists of nine matrices; two of them correspond to drug-target interaction, two others correspond to drug-disease associations, two others are correspond to disease-target associations. Three remaining are drug-drug, disease-disease and target-target relations as separate matrices. We convert these nine matrices to six by merging of matrices related to similar concept (for example two drug-target interaction matrices are merged to produce one matrix and similarly for others). Final interaction prediction is achieved by sorting the rows of these matrices.

Algorithm 1 is implemented in C# using Visual Studio 2013 (the source code is available upon request from the author).

### 3.6 Key points in Heter-LP

The main idea of our projection phase is extracted from [60] and [61]. DT-Hybrid algorithm [22] which is one of the best approaches for prediction of drug-target interactions, is also a recommendation method based on projection. DT-Hybrid attempts to provide a similarity matrix for a set of vertices (like Y) by using the similarity between other sets of vertices of the corresponding bipartite network (like X) across their relations. However, here we focus on relationships between two sets and try to extract a topological similarity matrix by it. In this way we reduce the required computational tasks in comparison to DT-Hybrid. Moreover, we use the similarity matrix of set X at label propagation phase. In this way, the similarity effects of vertices are not only through direct links but also we provide a kind of transitivity.

In label propagation phase, although there are some similarities with MINProp algorithm [59] there are several notable advantages. First, we use integrated data from primiary input data and output of the projection phase as input for label propagation section. Second, there is one less iterative loop here which reduce the computational complexity of the algorithm impressively. It is noticeable that other heterogeneous label propagation algorithms like LPMIHN [32] also could not reduce the iterative loops in this way.

Moreover, Heter-LP has some more general advantages which include:

- No need to any inadvisable preprocessing of data. In other heterogeneous label propagation algorithms, it is assumed that there is coincidence between one homogenous set of vertices with equivalent vertices in heterogeneous networks. To provide it, they had to remove some informative data in the preprocessing phase. Heter-LP does not need to such a preprocessing because of no need to such a coincidence.
- No need to know the negative samples. Here negative samples mean interactions which could not exist biologically. Most of other drug repositioning methods need to know negative samples and try to provide them randomly (there is no category for negative samples in this field).
- Heter-LP can predict interactions of *new* drugs (where a drug has no known target) and *new* targets (where a target has no known drug). This property is considered only in a few other methods.
- The other benefit of the proposed method is its ability to predict both trivial and non-trivial interactions. Trivial means the interactions which are predictable at first glance by everyone because of the existing similarities. Non-trivial interactions are the interactions that more evidence is required to find them.

## 4 Convergence Augment

An important part of the proposed method is the "LabelPropagation($f_1$, $f_2$, $f_3$)" function. This function is based on an iterative algorithm whose convergence is here demonstrated. In fact, we will show that the sequences *{$f_1(t)$}, {$f_2(t)$}, and {$f_3(t)$}* will ultimately converge and their corresponding answers are:

$$f_1^* = (I - \alpha M_1)^{-1}[(1-\alpha)^2 y_1 + (1-\alpha)^3 \alpha S_{1,2} y_2 + (1-\alpha)^3 \alpha S_{1,3} y_3]$$

$$f_2^* = (I - \alpha M_2)^{-1}[(1-\alpha)^2 y_2 + (1-\alpha)^3 \alpha S_{2,1} y_2 + (1-\alpha)^3 \alpha S_{2,3} y_3]$$

(3)

$$f_3^* = (I - \alpha M_3)^{-1}[(1-\alpha)^2 y_3 + (1-\alpha)^3 \alpha S_{3,1} y_1 + (1-\alpha)^3 \alpha S_{3,2} y_2]$$

Without loss of generality, we consider the same $\alpha$ for all sub-networks and rewrite $(1-\alpha)$ as $\beta$. So our first iterative equations will be as below:

$$\begin{aligned}
f_1(t) &= \beta(\beta y_1 + \alpha S_{1,2} f_2(t-1) + \alpha S_{1,3} f_3(t-1) + \alpha M_1 f_1(t-1)) \\
f_2(t) &= \beta(\beta y_2 + \alpha S_{2,1} f_1(t) + \alpha S_{2,3} f_3(t-1) + \alpha M_2 f_2(t-1)) \\
f_3(t) &= \beta(\beta y_3 + \alpha S_{3,1} f_1(t) + \alpha S_{3,2} f_2(t) + \alpha M_3 f_3(t-1))
\end{aligned} \quad (4)$$

By substitution of the above, we find:

$$\begin{aligned}
f_1(t) &= \left[\beta^2 \sum_{i=0}^{t-1} (\alpha M_1)^i y_1 + (\alpha M_1)^t y_1\right] \\
&+ \left[\beta^3 \alpha \sum_{i=0}^{t-2} (\alpha M_1)^i S_{1,2} y_2 + \beta \alpha (\alpha M_1)^{t-1} S_{1,2} y_2\right] \\
&+ [\beta^3 \alpha \sum_{i=0}^{t-2} (\alpha M_1)^i S_{1,3} y_3 + \beta \alpha (\alpha M_1)^{t-1} S_{1,3} y_3] + P_1 \\
f_2(t) &= \beta^2 \sum_{i=0}^{t} (\alpha M_2)^i y_2 + \beta^3 \alpha \sum_{i=0}^{t-1} (\alpha M_2)^i S_{2,1} y_1 + \beta^3 \alpha \sum_{i=0}^{t-1} (\alpha M_2)^i S_{2,3} y_3 + P_2 \\
f_3(t) &= \beta^2 \sum_{i=0}^{t} (\alpha M_3)^i y_3 + \beta^3 \alpha \sum_{i=0}^{t-1} (\alpha M_3)^i S_{3,1} y_1 + \beta^3 \alpha \sum_{i=0}^{t-1} (\alpha M_3)^i S_{3,2} y_2 + P_3
\end{aligned} \quad (5)$$

Where, each $P_i$ is a summation of different $t$ power of $S_i$ and $S_{i,j}$ and they will converge to zero as iterations progress.

The final results will be achieved by $\lim_{t \to \infty} f_j(t)$ which are:

$$f_1^* = \lim_{t \to \infty} f_1(t) = \lim_{t \to \infty}\left[\beta^2 \sum_{i=0}^{t-1}(\alpha M_1)^i y_1 + (\alpha M_1)^t y_1\right]$$

$$+ \lim_{t \to \infty}\left[\beta^3 \alpha \sum_{i=0}^{t-2}(\alpha M_1)^i S_{1,2}\, y_2 + \beta\alpha(\alpha M_1)^{t-1}S_{1,2}\, y_2\right]$$

$$+ \lim_{t \to \infty}\left[\beta^3 \alpha \sum_{i=0}^{t-2}(\alpha M_1)^i S_{1,3}\, y_3 + \beta\alpha(\alpha M_1)^{t-1}S_{1,3}y_3\right] + \lim_{t \to \infty} P_1 \quad (6)$$

$$\cong \lim_{t \to \infty}\left[\beta^2 \sum_{i=0}^{t}(\alpha M_1)^i y_1\right] + \lim_{t \to \infty}\left[\beta^3 \alpha \sum_{i=0}^{t-1}(\alpha M_1)^i S_{1,2}\, y_2\right]$$

$$+ \lim_{t \to \infty}\left[\beta^3 \alpha \sum_{i=0}^{t-1}(\alpha M_1)^i S_{1,3}\, y_3\right] + \lim_{t \to \infty} P_1$$

We know that:

$$\lim_{t \to \infty}\left[\beta^2 \sum_{i=0}^{t}(\alpha M_1)^i y_1\right] = \beta^2 (I - \alpha M_1)^{-1} y_1$$

$$\lim_{t \to \infty}\left[\beta^3 \alpha \sum_{i=0}^{t-1}(\alpha M_1)^i S_{1,2}\, y_2\right] = \beta^3 \alpha (I - \alpha M_1)^{-1} S_{1,2} y_2$$

$$\lim_{t \to \infty}\left[\beta^3 \alpha \sum_{i=0}^{t-1}(\alpha M_1)^i S_{1,3}\, y_3\right] = \beta^3 \alpha (I - \alpha M_1)^{-1} S_{1,3}\, y_3 \quad (7)$$

$$\lim_{t \to \infty} P_1 = 0$$

So the final equation for $f_1^*$ will be:

$$f_1^* = \lim_{t \to \infty} f_1(t) = (1-\alpha)^2 (I - \alpha M_1)^{-1} y_1 + (1-\alpha)^3 \alpha (I - \alpha M_1)^{-1} S_{1,2} y_2 +$$
$$(1-\alpha)^3 \alpha (I - \alpha M_1)^{-1} S_{1,3} y_3 = (I - \alpha M_1)^{-1}[(1-\alpha)^2 y_1 + (1-\alpha)^3 \alpha S_{1,2} y_2 + \quad (8)$$
$$(1-\alpha)^3 \alpha S_{1,3} y_3]$$

In the same way for $f_2^*$ and $f_3^*$, we can write:

$$f_2^* = \lim_{t \to \infty} f_2(t) = (1-\alpha)^2 (I - \alpha M_2)^{-1} y_2 + (1-\alpha)^3 \alpha (I - \alpha M_2)^{-1} S_{2,1} y_1$$
$$+ (1-\alpha)^3 \alpha (I - \alpha M_2)^{-1} S_{2,3} y_3 \quad (9)$$
$$= (I - \alpha M_2)^{-1}[(1-\alpha)^2 y_2 + (1-\alpha)^3 \alpha S_{2,1} y_2 + (1-\alpha)^3 \alpha S_{2,3} y_3]$$

$$f_3^* = \lim_{t\to\infty} f_3(t) = (1-\alpha)^2(I-\alpha M_3)^{-1}y_3 + (1-\alpha)^3\alpha(I-\alpha M_3)^{-1}S_{3,1}y_1$$
$$+ (1-\alpha)^3\alpha(I-\alpha M_1)^{-1}S_{3,2}y_2$$
$$= (I-\alpha M_3)^{-1}[(1-\alpha)^2 y_3 + (1-\alpha)^3\alpha S_{3,1}y_1$$
$$+ (1-\alpha)^3\alpha S_{3,2}y_2]$$

We express the resulting equation in closed-form as Eq. (10) below:

$$f_i^* = (I-\alpha M_i)^{-1}[(1-\alpha)^2 y_i + (1-\alpha)^3\alpha \sum_{j\neq i} S_{i,j}y_j] \ \ for\ i,j = 1,2,3 \quad (10)$$

Here we have three homogeneous sub-networks. It can be easily verified that if we set *i,j = 1,2,…, k*, then Eq. (10) will be correct for a heterogeneous network with *k* different homogenous sub-networks.

Now we must choose a value for the constant parameter α. It seems that α cannot be arbitrarily large. If we let *α→0*, then all final labels will be equal to the initial ones. If we increase *α* from zero, we will come to a point at which $(I-\alpha M_i)^{-1}$ will diverges and cause the divergence of $f_i$. This will be happen when the determinant of $(I-\alpha M_i)$ passes through zero [62]. We can rewrite this condition as $det(M_i - \alpha^{-1}I) = 0$. This will be happen when the roots of $\alpha^{-1}$ are equal to the eigenvalues of $M_i$. When $\alpha^{-1}$ becomes equal to the largest eigenvalue ($k_i$) of $M_i$, the determinant first crosses zero. Therefore, we must choose a value of α less than *1/$k_i$*.

The final labels can be calculated directly from Eq. (10). However, this would require the inversion of $(I-\alpha M_i)$. In order to invert this *n\*n* matrix, a time proportional to $O(n^3)$ is required by matrix multiplication algorithms, which can at best be reduced to $O(n^{2.373})$ through algorithmic optimization[12]. Therefore, we instead prefer to use the direct equations represented in Eq. (4). By repeating the process several times, the results will converge to the correct values.

We here measure the required runtime for each step (projection and label propagation) separately. In the projection phase, we need to know the degrees of all the vertices of the interaction matrix. The required time is $O(\ 2(|V_i||V_j|)$. The expected time to project one interaction matrix on one similarity matrix is $O(\ |V_i|^2|V_j|)$ which is repeated six times. So the total runtime for projection phase is $O(6\ (\ 2(|V_i||V_j| + |V_i|^2|V_j|))$.

---

[12] https://en.wikipedia.org/wiki/Computational_complexity_of_mathematical_operations

For label propagation on each homogeneous sub-network, the required time is equal to $O(t(|V_i| + \sum_{j \neq i}|V_i||V_j| + |V_i|^2))$ and the total runtime of label propagation phase will be:

$$O(t \sum_{i=1}^{3}(|V_i| + \sum_{j \neq i}|V_i||V_j| + |V_i|^2))$$

The total time for calculation is proportional to *t*, where *t* is the number of iterations required to reach convergence. The value of *t* depends on the input data structures and the parameter α. Therefore, we cannot determine the necessary time independently on the runtime parameters. However, in our experiments, α was always smaller than and equal to 0.3 and the value of *t* was always smaller than 10.

## 5   Regularization Framework

Here we develop a regularization framework for the proposed label propagation algorithm. First, an objective function is determined. Then we will show that this function is strictly convex and will therefore have a globally optimal solution. We describe this solution and find it equal to the results of the previous section. In this way, it is proved that our proposed method will find an optimal solution.

**The objective function:** A a cost function is defined to propagate a label in a heterogeneous network *G (V, E)*, as Eq. (11) below:

$$\Omega(f) = \sum_{i=1}^{3}\left(f_i^T \Delta^{(i)} f_i + \mu_i \|f_i - y_i\|^2 + \mu_i \sum_{i=1}^{2}\sum_{j=2}^{3}[f_i^T f_j^T]\sum^{(i,j)}\begin{bmatrix}f_i\\f_j\end{bmatrix}\right) \quad (11)$$

where

- $f \in R^{|v|}$ is label vector,
- $y_i$ is the initial label vector
- $\Delta^{(i)} = I - M_i$
- $\Sigma^{(i,j)} = I - \begin{bmatrix} 0 & S_{i,j} \\ (S_{i,j})^T & 0 \end{bmatrix}$ is the normalized graph laplacian of $S^{i,j}$
- and $\|.\|$ indicates a vector norm.

$\Omega(f)$ consists of three cost terms:

- $f_i^T \Delta^{(i)} f_i$ is a smoothness term on the homogenous sub-network $G^{(i)} = (V^{(i)}, E^{(i)})$ which causes the similarity of labels of the connected nodes in the network $G^{(i)}$.
- $\|f_i - y_i\|^2$ is a fitting term which tends to keep the new label values near to the initial ones.

- $[f_i^T f_j^T] \Sigma^{(i,j)} \begin{bmatrix} f_i \\ f_j \end{bmatrix}$ is a smoothness term on the heterogeneous sub-network $G^{(i,j)} = (V^{(i)} \cup V^j, E^{(i,j)})$ which ensures the similarity of labels of connected nodes in the network $G^{(i,j)}$.

**Proposition 1:** $\Omega(f)$ is strictly convex.

*Proof:* All of $\Delta^i$ and $\Sigma^{(i,j)}$ terms are graph laplacian. This means that they are positive semi-definite and cause the convexity of the first and third cost terms of $\Omega(f)$. The second cost term is also convex. All of them are multiplied by some positive constants. So $\Omega(f)$ is a nonnegative-weighted sum of some convex function and is therefore itself a convex function [63].

On the other hand, the Hessian matrix of $\Omega(f)$ is a summation of $\Delta^i$, $\Sigma^{(i,j)}$ and $I$. The $\Delta^i$ and $\Sigma^{(i,j)}$ terms are positive semi-definite and $I$ is positive definite, so the Hessian matrix of $\Omega(f)$ is positive definite[13].

Therefore, $\Omega(f)$ is strictly convex. ∎

**Proposition 2:** The optimal solution of $\Omega(f)$ is:

$$f_i^* = (I - \alpha 1_i M_i)^{-1}[\alpha 2_i y_i + \alpha 3_i \sum_{j \neq i} S_{i,j} f_j] \tag{12}$$

*Proof:* As proven in Proposition 1, $\Omega(f)$ is strictly convex, so it can be solved by alternating optimization [63]. For each $f_i$, all $f_j$ terms for $j \in \{n | n \in \mathbb{N}, 1 \leq n \leq 3, n \neq i\}$ are considered constant. We can then differentiate $\Omega(f)$ with respect to $f_i$ and set it equal to zero to compute the solution:

$$d(\frac{\Omega(f)}{f_i}) = 0$$

$$f_i^* - M_i f_i^* + 2\mu_i(f_i^* - y_i) + 2f_i^* - \sum_{j \neq i} \mu_i S^{i,j} f_j = 0 \tag{13}$$

We will have:

$$((3 + 2\mu_i)I_i - M_i)f_i^* - 2\mu_i y_i - \sum_{j \neq i} \mu_i S^{i,j} f_j = 0$$

$$f_i^* = \left(I - \frac{M_i}{3 + 2\mu_i}\right)^{-1} \left(\frac{2\mu_i}{3 + 2\mu_i} y_i + \frac{\mu_i}{3 + 2\mu_i} \sum_{j \neq i} S^{i,j} f_j\right) \tag{14}$$

---

[13] If $\nabla^2 f(x) \succ 0$ for all $x \in$ domain(f), then f is strictly convex 63 Boyd, S., and Vandenberghe, L.: 'Convex Optimization' (Cambridge University Press, 2004. 2004).

We will set:

$$\alpha 1_i = \frac{1}{3+2\mu_i}, \quad \alpha 2_i = \frac{2\mu_i}{3+2\mu_i}, \quad \alpha 3_i = \frac{\mu_i}{3+2\mu_i} \tag{15}$$

So the closed form of the solution will be:

$$f_i^* = (I_i - \alpha 1_i M_i)^{-1}[\alpha 2_i y_i + \alpha 3_i \sum_{j \neq i} S^{(i,j)} f_j] \blacksquare \tag{16}$$

**Proposition 3:** The proposed algorithm will minimize the objective function $\Omega(f)$.

Proof: We should show that the result of Proposition 2 is equal with our answer in Eq. (10).

We could rewrite the iterative Eq. (3) as:

$$f_i(t) = (1-\alpha)^2 y_i + \alpha M_i f_i(t-1) + \sum_{j \neq i}(1-\alpha)^2 \alpha S^{i,j} f_j(t-1)$$
$$i, j = 1,2,3 \tag{17}$$

This is the equation solved by our proposed method and the result is:

$$f_i^* = (I - \alpha M_i)^{-1}[(1-\alpha)^2 y_i + (1-\alpha)^3 \alpha \sum_{j \neq i} S^{i,j} y_j] \tag{18}$$

If we set:
$\alpha 1_i = \alpha, \quad \alpha 2_i = (1-\alpha)^2, \quad \alpha 3_i = (1-\alpha)^3 \alpha$
Both Eq. (16) and Eq. (18) will be equal. $\blacksquare$

So we have an optimization problem with the objective function represented by Eq. (11) which is strictly convex and, therefore, the proposed method can find its global minimum.

# 6 Performance evaluation

In this section, we report on the results of our experiments to evaluate the performance of the proposed algorithm using the datasets described in Section 3.

## 6.1 Statistical analysis

We considered three different scores as indicators for prediction accuracy:

1. Area Under the Curve (AUC) of the Receiver Operating Characteristics (ROC) curve
2. Area Under the Precision-Recall (AUPR) curve
3. BestAccuracy

ROC is the plot of true positive rate (TPR) as a function of false positive rate (FPR) evaluated at various decision thresholds, where the true positives are correctly predicted true interactions and the false positives are predicted interactions not present in the gold standard set of interactions.

Precision is defined as the fraction of true drug targets identified among the candidate proteins ranked above the particular decision threshold. Recall is the fraction of true drug targets identified from among the total number of true drug targets in the gold standard set of interactions.

Although AUC represents the overall performance of the algorithm, previous studies have demonstrated that precision-recall curves more accurately assess a method's performance in the face of skewed node degree distributions in scale-free biological networks [9]. Precision-recall (PR) curves are also more informative when significant class imbalance exists. Furthermore, a curve dominates in ROC space if and only if it also dominates in PR space [35].

Accuracy measures the difference between a measurement with the actual value. It could be defined as equation (19):

$$accuracy = \frac{TP+TN}{TP+FP+TN+FN} \quad (19)$$

Here, TP = True Positives, FP = False Positives, TN = True Negatives, and FN = False Negatives.

The highest achieved accuracy repeated experiments with the same parameter values over the same data is defined as "BestAccuarcy".

**6.1.1 Results based on gold standard dataset**

We performed 10-fold cross-validation (10-CV) to analyze the performance of the proposed Heter-LP method. In a 10-CV experiment, we split the dataset into ten subsets of equal size; each subset is then taken in turn as a test set, and the training is performed on the remaining nine sets.

Table 1 shows the results of the proposed method during 10-CV over the gold standard datasets of Yamanishi 2008 [24] with augmented information as described in section 3.2.1. One can observe that the results of drug-target prediction are the strongest. This indicates that Heter-LP can predict drug-target interactions more accurately than the other two interactions. We repeated this test using some other methods and observed the same trend. This result appears to be due to incompleteness of the input data. As explained before, the available golden standard dataset doesn't contain any information about diseases and our attempt for the addition of such information was insufficient to create a suitable complete dataset. In other words, the disease similarity matrix and the primitive interaction matrices used here are not sufficiently informative.

**6.1.2 Comparison with state of the art methods on gold standard data**

To provide the opportunity to compare the performance of different methods, two-column charts are represented in Figure 3 and Figure 4. Represented performances are self-reported by the corresponding papers (more explanation is represented in supplementary material). The corresponding numerical values are available in Table 3 of supplementary materials. We found that "DT-Hybrid" performs best. So we downloaded this package and evaluated it for two reasons. Firstly, to analyze two other types of interactions mentioned before (disease-target, drug-disease) and secondly, to find the accurate performance of this package with 10-CV.

**6.1.3 Comparison of 10-fold CV results**

In total, DT-Hybrid's results are weaker than our proposed method in disease-target and drug-disease prediction. The results are presented in Table 2 and Table 3. It is difficult to determine which method is best overall; in 8 of 12 cases Heter-LP is the top-performing methods while in 4 of 12 cases, DT-Hybrid is the best.

We also implemented an example of a poorly performing method, MINProp [59], and evaluated it using the gold standard data. The 10-CV results are represented in Table 2 and Table 3. As you can see, they are substantially weaker than the two other methods.

Table 1 Results of Heter-LP method during 10-CV testing on the gold standard dataset

| Interaction | Dataset: E | | | Dataset: GPCR | | | Dataset: IC | | | Dataset: NR | | |
|---|---|---|---|---|---|---|---|---|---|---|---|---|
| | *AUC* | *AUP* | *Best* | *AUC* | *AUP* | *Best* | *AUC* | *AUP* | *Best* | *AUC* | *AUP* | *Best* |

|  |  | R | Acc |  | R | Acc |  | R | Acc |  | R | Acc |
|---|---|---|---|---|---|---|---|---|---|---|---|---|
| *Drug-Disease* | 0.8292 | 0.8475 | 0.7231 | 0.8606 | 0.8697 | 0.7468 | 0.8400 | 0.8539 | 0.7340 | 0.8640 | 0.8625 | 0.7582 |
| *Drug-Target* | 0.9918 | 0.4967 | 0.9917 | 0.9928 | 0.8575 | 0.9873 | 0.9878 | 0.7684 | 0.9856 | 0.9823 | 0.8965 | 0.9789 |
| *Disease-Target* | 0.9147 | 0.8020 | 0.9449 | 0.8529 | 0.7381 | 0.8850 | 0.9163 | 0.8096 | 0.9454 | 0.9383 | 0.9459 | 0.9444 |

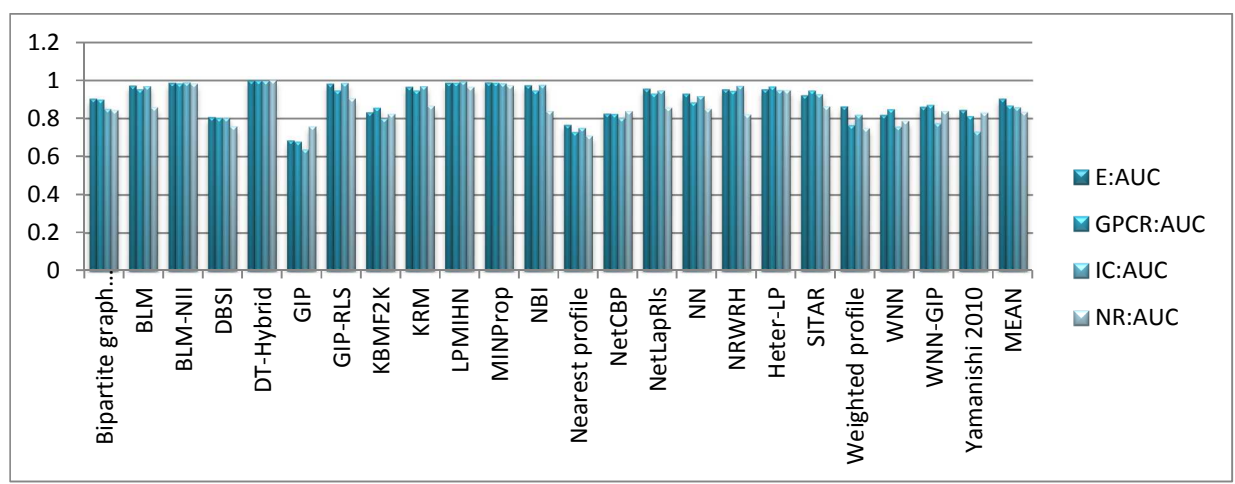

**Figure 3 Self-reported AUC of various methods in gold standard datasets.**

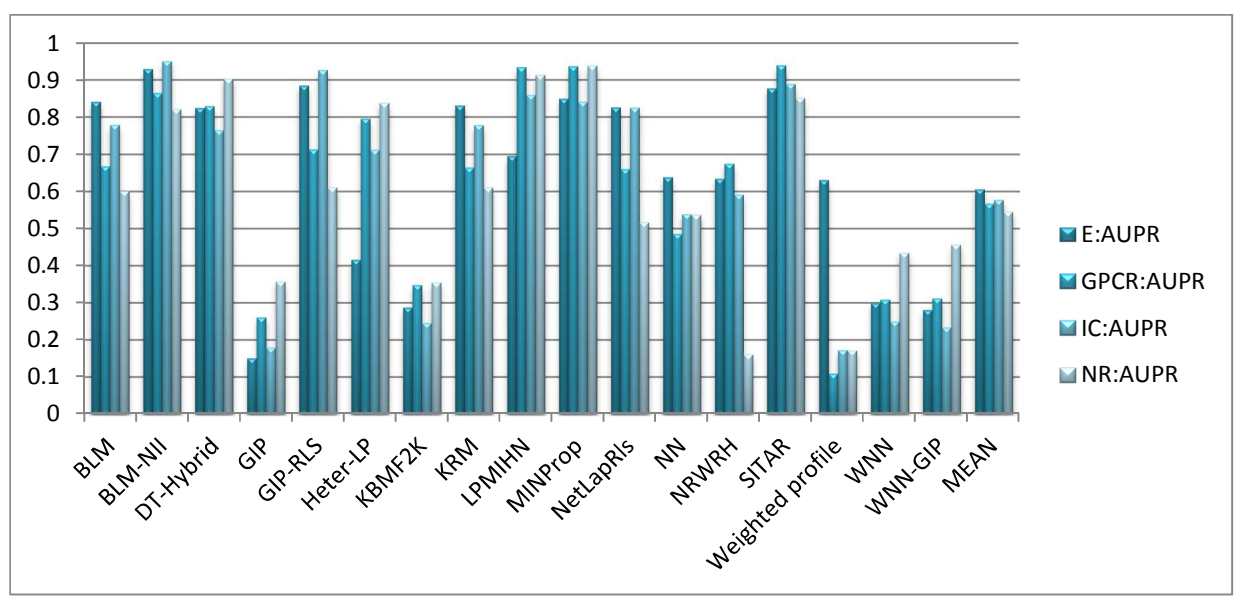

**Figure 4 Self-reported AUPR of some different methods in gold standard datasets.**

Table 2 Results of 10-fold cross validation on E & GPCR datasets.

| Method | Interaction | Dataset: E | | | Dataset: GPCR | | |
|---|---|---|---|---|---|---|---|
| | | *AUC* | *AUPR* | *BestACC* | *AUC* | *AUPR* | *BestACC* |
| *DT-Hybrid (R)* | *Drug-disease* | 0.715656 | 0.736074 | 0.695336 | 0.695598 | 0.713113 | 0.667188 |
| | *Drug-target* | 0.947071 | **0.824319** | **0.996805** | 0.937225 | 0.829404 | **0.987817** |
| | *Disease-target* | 0.637279 | 0.641289 | 0.612389 | 0.668512 | 0.682871 | 0.636683 |
| *Heter-LP* | *Drug-disease* | 0.77056 | 0.81442 | 0.70544 | 0.79321 | **0.83176** | 0.72315 |
| | *Drug-target* | **0.95262** | 0.41505 | 0.99142 | **0.96747** | 0.79579 | 0.98584 |
| | *Disease-target* | 0.84988 | 0.75554 | 0.88614 | 0.79582 | 0.69718 | 0.84017 |
| *MINProp* | *Drug-disease* | 0.4776 | 0.4745 | 0.5039 | 0.5 | 0.2479 | 0.5043 |
| | *Drug-target* | 0.5108 | 0.0119 | 0.9901 | 0.4978 | 0.0239 | 0.9700 |
| | *Disease-target* | 0.5195 | 0.5122 | 0.5222 | 0.5014 | 0.5326 | 0.5033 |

Table 3 Results of 10-fold cross validation on IC & NR datasets.

| Method | Interaction | Dataset: IC | | | Dataset: NR | | |
|---|---|---|---|---|---|---|---|
| | | *AUC* | *AUPR* | *BestACC* | *AUC* | *AUPR* | *BestACC* |
| *DT-Hybrid (R)* | *Drug-disease* | 0.673724 | 0.707504 | 0.641833 | 0.689028 | 0.698906 | 0.661216 |
| | *Drug-target* | 0.923267 | 0.764408 | 0.983021 | 0.947913 | 0.903397 | **0.987536** |
| | *Disease-target* | 0.648534 | 0.655571 | 0.627719 | 0.757061 | 0.771653 | 0.725762 |
| *Heter-LP* | *Drug-disease* | 0.77764 | **0.81822** | 0.71237 | 0.80473 | 0.82928 | 0.73516 |
| | *Drug-target* | **0.94828** | 0.71143 | **0.98364** | 0.94797 | 0.83816 | 0.97523 |
| | *Disease-target* | 0.85669 | 0.77125 | 0.89916 | 0.88147 | **0.90672** | 0.89813 |
| *MINProp* | *Drug-disease* | 0.5 | 0.2497 | 0.5005 | 0.5 | 0.2470 | 0.5059 |
| | *Drug-target* | 0.5 | 0.0172 | 0.9655 | 0.5229 | 0.0679 | 0.9359 |
| | *Disease-target* | 0.5 | 0.2498 | 0.5004 | 0.4928 | 0.4726 | 0.5145 |

#### 6.1.4   Prediction of a deleted interaction

A standard approach for validating a method is to remove some of the desirable entries from the input, then to run the process and finding the results (as one does during a cross-validation experiment). In this regard, we performed two distinct examinations. We chose an arbitrary drug (or target) and, as Test1, we deleted one of its interactions from the input dataset and, as Test2, we deleted all of its interactions with targets (or drugs) from the dataset. Test1 is explained here and Test2 will be described in the next section.

By deleting a single interaction, we investigate the ability of the method in predicting new interactions for known drugs or targets (i.e. nodes that have some known interaction with others). We perform different analysis tasks in this regard, and here we illustrate a case study. D00232 is a drug with three interaction with targets: hsa:1128, hsa:1129 and hsa:1131. We deleted its interaction with hsa:1129 from the input network and investigated the results. Both Heter-LP and DT-Hybrid correctly predict this removed interaction, as can be seen in Table 4. This list also contains several high scoring novel drug-target predictions which merit experimental validation.

**Table 4 Top 20 predicted targets of Drug: D00232 by proposed method**

| NO. | Heter-LP | DT-Hybrid |
|---|---|---|
| 1 | **hsa:1128** | **hsa:1128** |
| 2 | **hsa:1131** | **hsa:1131** |
| 3 | **hsa:1129** | **hsa:1129** |
| 4 | hsa:11255 | hsa:1132 |
| 5 | hsa:154 | hsa:3269 |
| 6 | hsa:3269 | hsa:1133 |
| 7 | hsa:153 | hsa:3360 |
| 8 | hsa:1813 | hsa:8843 |
| 9 | hsa:148 | hsa:1813 |
| 10 | hsa:4988 | hsa:3356 |
| 11 | hsa:185 | hsa:148 |
| 12 | hsa:150 | hsa:3358 |
| 13 | hsa:3577 | hsa:1812 |
| 14 | hsa:3274 | hsa:1815 |
| 15 | hsa:152 | hsa:146 |
| 16 | hsa:147 | hsa:147 |
| 17 | hsa:3360 | hsa:59340 |
| 18 | hsa:146 | hsa:11255 |
| 19 | hsa:1814 | hsa:150 |
| 20 | hsa:155 | hsa:151 |

**6.1.5    Prediction of pseudo-new drugs**

As mentioned above, our method can predict the interactions between new targets and new drugs correctly. This feature offers a great advantage that most of the existing methods do not provide. In this regard, we perform Test2 to create a pseudo-new drug and compare the results of our proposed method with DT-Hybrid.

In the previous section, we explained that drug D00232 has three targets in the gold standard drug-target interaction. We first deleted all of these interactions. In this way, D00232 is like a new drug in that it no longer has any known interactions with any target in the input network. Both Heter-LP and DT-Hybrid were then applied to the censored dataset. The top 20 targets predicted by Heter-LP are presented in Table 5. All of the desired targets (which we deleted from the input data) are predicted successfully by our method (see bold entries in Table 5). Recall that these were recovered from among 989 possible targets. However, DT-Hybrid could not predict any target for this new drug (all of the entries of D00232 in its output were zero).

Similar examinations were conducted for some other drugs (like D05353, D00227) and for a variety of creating pseudo-new targets by removing all known interactions with drugs. Results were largely consistent with those of Test2, but results are excluded due to space limitation.

Table 5 Top 20 predicted targets of Drug: D00232 by proposed method

| NO. | Predicted target |
|---|---|
| 1 | hsa:154 |
| 2 | hsa:3269 |
| 3 | hsa:153 |
| **4** | **hsa:1128** |
| 5 | hsa:1813 |
| 6 | hsa:148 |
| 7 | hsa:4988 |
| 8 | hsa:185 |
| **9** | **hsa:1129** |
| 10 | hsa:3577 |
| 11 | hsa:150 |
| 12 | hsa:3274 |
| 13 | hsa:1814 |
| 14 | hsa:146 |
| 15 | hsa:3360 |
| **16** | **hsa:1131** |

| 17 | hsa:3356 |
| 18 | hsa:147 |
| 19 | hsa:155 |
| 20 | hsa:151 |

## 6.2 Experimental analysis

Statistical analysis confirms the ability of the proposed method in predicting potential interactions, now is the time to investigate its practical effectiveness. In this regard, we used all the data as training set and examined new predicted interactions. Then we rank the unknown interactions on their scores and extract the top 20 predictions. These novel interactions were checked manually using the online version of DrugBank[14], Supertarget[15], KEGG Drug[16] and Therapeutic Target Database (TTD)[17].

This test is performed two times, one based on the gold standard dataset and the other one using the introduced independent datasets (Section 3.2.2). Because of space limitations, the full predicted lists are placed in supplementary materials.

We categorized the predictions in two groups, trivial and non-trivial ones. Trivial predictions could be predicted by straightforward and primary investigations of the input data. Non-trivial could not be quickly discovered from input data. It seems that an effective method should be able to identify both types sufficiently. Some examples are represented in the following sections.

### 6.2.1 Experimental analysis based on gold standard dataset

We sorted the predicted list of unknown drug-target interactions of each group (E, GPCR, IC, NR) and extracted the top 20 ones of each group separately. Because of space limitations, only the results of GPCR are represented here in Table 6; others are available in Tables 4-6 of supplementary materials. A similar investigation was done by some of the other methods like BLM [30], KBMF2k [23], LapRLS [33], LPMIHN [32], NetCBP [28], NRWRH [48], RLS-Kron[18] [30], WNN [31]. Table 7 of supplementary materials is a brief comparison of the results of experimental analysis of different methods.

Verified predictions in Table 6 are denoted by the name of the related source. As you can see, seven of 20 GPCR interactions are verified. Furthermore, a number of the non-validated predictions have additional supporting biological evidence that are out the scope of this paper. One non-trivial example prediction is discussed in the following case study.

---

[14] http://www.drugbank.ca/
[15] http://insilico.charite.de/supertarget/
[16] http://www.genome.jp/kegg/drug/
[17] http://bidd.nus.edu.sg/group/cjttd/
[18] Regularized Least Squares (RLS) with Kronecker sum kernel

**Table 6 The top 20 new predicted interactions in GPCR dataset**

| NO. | pair | | Annotation | | | Verification source |
|---|---|---|---|---|---|---|
| | Drug | Target | Drug | Target | UniprotName of target | |
| 1 | D00542 | hsa:338442 | Halothane (JP17/USP/INN) | G-protein coupled receptor 109A | NIAR1_HUMAN | |
| 2 | D02358 | hsa:154 | Metoprolol (USAN/INN) | Beta-2 adrenergic receptor | ADRB2_HUMAN | SuperTarget |
| 3 | D04625 | hsa:154 | Isoetharine (USP) Isoetarine (INN) | Beta-2 adrenergic receptor | ADRB2_HUMAN | KEGG |
| 4 | D02614 | hsa:154 | Denopamine (JAN/INN) | Beta-2 adrenergic receptor | ADRB2_HUMAN | |
| 5 | D02147 | hsa:153 | Albuterol (USP) Salbutamol | Beta-1 adrenergic receptor | ADRB1_HUMAN | SuperTarget |
| 6 | D02359 | hsa:153 | Ritodrine (USAN/INN) | Beta-1 adrenergic receptor | ADRB1_HUMAN | |
| 7 | D00683 | hsa:153 | Albuterol sulfate (USP) Salbutamol sulfate (JP17) | Beta-1 adrenergic receptor | ADRB1_HUMAN | SuperTarget |
| 8 | D05792 | hsa:153 | Salmeterol (USAN/INN) | Beta-1 adrenergic receptor | ADRB1_HUMAN | |
| 9 | D00688 | hsa:153 | Terbutaline sulfate (JP17/USP) | Beta-1 adrenergic receptor | ADRB1_HUMAN | SuperTarget |
| 10 | D00684 | hsa:153 | Bitolterol mesylate (USAN) Bitolterol mesilate (JAN) | Beta-1 adrenergic receptor | ADRB1_HUMAN | |
| 11 | D01386 | hsa:153 | Ephedrine hydrochloride (JP17/USP) | Beta-1 adrenergic receptor | ADRB1_HUMAN | KEGG |
| 12 | D00687 | hsa:153 | Salmeterol xinafoate (JAN/USAN) | Beta-1 adrenergic receptor | ADRB1_HUMAN | SuperTarget |
| 13 | D00673 | hsa:3269 | Ranitidine hydrochloride (JP17/USP) | Histamine H1 receptor | HRH1_HUMAN | |
| 14 | D03503 | hsa:3269 | Cimetidine hydrochloride (USP) | Histamine H1 receptor | HRH1_HUMAN | |
| 15 | D00422 | hsa:3269 | Ranitidine (USAN/INN) | Histamine H1 receptor | HRH1_HUMAN | |
| 16 | D00440 | hsa:3269 | Nizatidine (JP17/USP/INN) | Histamine H1 receptor | HRH1_HUMAN | |

| | | | | | |
|---|---|---|---|---|---|
| 17 | D00295 | hsa:3269 | Cimetidine (JP17/USP/INN) | Histamine H1 receptor | HRH1_HUMAN |
| 18 | D00765 | hsa:1128 | Rocuronium bromide (JAN/USAN/INN) | Muscarinic acetylcholine receptor M1 | ACM1_HUMAN |
| 19 | D01346 | hsa:3269 | Bentiromide (JAN/USAN/INN) | Histamine H1 receptor | HRH1_HUMAN |
| 20 | D00760 | hsa:1128 | Doxacurium chloride (USAN/INN) | Rocuronium bromide (JAN/USAN/INN) | Muscarinic acetylcholine receptor M1 |

Although we also had predicted drug-disease and disease-target interactions, we do not discuss them here since none of the compared methods are capable of making such predictions. Instead, these interactions will be discussed in the experimental analysis based on independent datasets.

On the other hand, we claim that our proposed method could predict trivial and non-trivial interactions. Here we will explain two case studies to demonstrate this claim.

*A trivial case study: (D02358 & hsa:154)*

D02358 is the KEGG id of drug Metoprolol (USAN/INN) and hsa:154 is the KEGG id of protein target Beta-2 adrenergic receptor which is also known by its UniProt name ADRB2_HUMAN. No interaction is defined between D02358 and hsa:154 in our using gold standard dataset as input (the corresponding entry in $G_{1,3}$ is zero).

We searched the SuperTarget website in August 2016 and find hsa:154 as a target of D02358. We therefore consider this to be a trivial prediction since it could have been predicted via simple research. According to the input similarity matrix of protein targets, hsa:154 is the most similar target to hsa:153. However, according to our input interaction matrix, hsa:153 is the sole target of D02358. It is reasonable to introduce the pair (D02358 & hsa:154) as one the most probable candidates (row 2 of Table 4) and new experimental research (listed in SuperTarget) indeed verifies this interaction.

*A non-trivial case study: (D00673 & hsa:3269)*

D00673 is the KEGG id of drug Ranitidine hydrochloride and hsa:3269 is the KEGG id of protein target Histamine H1 receptor (UniProt name HRH1_HUMAN). No interaction is defined between D00673 and hsa:3269 in our using gold standard input dataset.

We first examined the input matrices in two ways to establish that this predicted interaction is non-trivial. First, we found the interacted targets of D00673 from the drug-target interaction input matrix then investigate their similar targets using the target similarity input matrix. Second, we found the interacted drugs of hsa:3269 from the drug-target interaction input matrix and identify similar drugs using the drug-drug similarity input matrix.

The only target of D00673 is hsa:3274. The most similar target to hsa:3274 is hsa:3360 and between 95 distinct targets, hsa:3269 ranks 37th in similarity to hsa:3274. Clearly, the predicted interaction (D00673 & hsa:3269) could not be predicted solely through target similarity.

We then found the drugs predicted to interact with hsa:3269 using the drug-target interaction input matrix. Its related drugs are:

D00234, D00283, D00300, D00364, D00454, D00480, D00493, D00494, D00520, D00521, D00665, D00666, D01242, D01295, D01324, D01332, D01713, D01717, D01782, D02327, D02354, D02361, D02566, D03621, D04979, D05129.

The most similar drug toD00673 is D00422. And from above mentioned drugs, the most similar one is D00480 which ranks 30$^{th}$ in terms of similarity with D00673. Clearly, the D00673 & hsa:3269 interaction could not have bene predicted based on drug similarity alone.

Now we will show that this prediction is plausible and should be considered as a good candidate for experimental validation. As verified by SuperTarget, D00673 now has 14 known targets, two of which are HRH2_HUMAN and HRH4_HUMAN. Our predicted target, HRH1_HUMAN, is highly similar to HRH4_HUMAN (SuperTarget data and DrugBank documentation clarify their similar aspects).

#### 6.2.2 Experimental analysis based on independent datasets

The gold standard data had value in that it enabled us to compare our proposed method with a wide variety of methods evaluated using the same data. However, the gold standard data is somewhat obsolete (2008) and incomplete (e.g. lacking drug-disease interaction data). We, therefore, created an updated and complete dataset to fully evaluate the capabilities of our proposed method, Heter-LP. These data were used as input of the proposed method and their results were analyzed. Here we here discuss one of its interesting predictions as a case study.

*Case study: Osteoarthritis with mild chondrodysplasia*

Osteoarthritis with mild chondrodysplasia is a type of skeletal disease due to the mutation of type II procollagen (COL2A1). It causes a progressive degeneration of the articular cartilage of joints with mild spinal chondrodysplasia[19,20].

---

[19] "Chondrodysplasia is a heterogeneous group of bone dysplasias, the common characteristic of which is stippling of the epiphyses in infancy." http://medical-dictionary.thefreedictionary.com/chondrodysplasia
[20] http://www.kegg.jp/kegg/disease/

Table 8 of supplementary materials presents its associated drugs and their corresponding targets (which we used as input data). The only known target of this disease is "hsa:1280" for which there is no known drug. The most similar disease (similarity higher than 0.3) to "Osteoarthritis with mild chondrodysplasia" and their associated drugs and targets are listed in Table 9 of supplementary materials.

We have predicted two new drugs for the treatment of this disease: Alendronate sodium, Alendronic acid. Alendronate sodium is a salted form of Alendronic acid. And, as expressed in DrugBank, is "for the treatment and prevention of osteoporosis in women and Paget's disease of bone in both men and women". Its treatment effects have two sides, one by its affinity for hydroxyapatite and the other, its inhibiting effect on FPP[21] synthase. Hydroxyapatite is part of the mineral matrix of bone and inhibition of FPP will inhibit osteoclast activity and reduce bone resorption[22].

In our input datasets both of Alendronate sodium and Alendronic acid are assigned for the treatment of Osteoporosis-pseudoglioma syndrome (OPPG), which is a skeletal disease "characterized by severe congenital osteoporosis with blindness"[23].

We assert that this prediction is plausible and merits further experimental validation because of the same category of "OPPG" and "Osteoarthritis with mild chondrodysplasia" and also the mechanism of action of Alendronic acid and Alendronate sodium. It is necessary to mention that this prediction is not a trivial one. As you can see in Table 9 of supplementary materials "OPPG" and "Osteoarthritis with mild chondrodysplasia" are not similar diseases in input dataset. Also, their input drugs and their input targets are not the same. The only target for "OPPG" in input datasets is hsa:4041 and, as mentioned before, the only known target of "Osteoarthritis with mild chondrodysplasia" is hsa:1280. To establish that this prediction could not have been made trivially on the basis of drug similarity, we provide a list of most similar drugs to "Alendronate" from input dataset in Table 10 of supplementary materials. Clearly, there is no similarity between the results in Table 8 and Table 10. No target similarity comparison is represented here because of the non-existence of hsa:4041 in input target similarity matrix.

## 7 Conclusion and discussion

Label propagation is an efficient technique to utilize both local and global features in a network for semi-supervised learning [59]. In spite of the growing interest in the use of heterogeneous networks in various scientific disciplines, there is insufficient attention being paid to label propagation on these networks. In this paper, we introduce a new label propagation algorithm on heterogeneous networks named Heter-LP. Its convergence to an optimal solution is discussed and proved. We had shown that there are fewer iteration loops in Heter-LP in comparison to other heterogeneous label propagation algorithms and the time complexity is acceptable.

---

[21] Farnesyl pyrophosphate
[22] http://www.drugbank.ca/drugs/DB00630
[23] http://www.kegg.jp/kegg/disease/

The Heter-LP algorithm was applied to the problem of drug repositioning to demonstrate its applicability. It was shown that Heter-LP can infer new interactions for disease-drug, drug-target, and disease-target relationships successfully through integrating heterogeneous information obtained from various types of resources at different levels of biological detail. In fact, we used both local and global features together by using label propagation. Furthermore, an advantage of the proposed model is that it does not require negative interactions for training, as experimental analysis rarely reports negative samples.

We provide a comprehensive statistical analysis of performance by using of 10-fold cross validation testing. The achieved AUC and AUPR outperform most existing state-of-the-art methods for drug repositioning. Although these parameters are weaker than some methods in some sense, our designed experimental analysis have proven some attractive abilities which there are in some rare practices. It is shown that Heter-LP could predict interactions of new drugs, targets and diseases correctly. Moreover, in spite of some methods which could predict only trivial interactions, Heter-LP could predict both trivial and non-trivial ones.

In total, the analysis demonstrates that label propagation is an effective algorithm to predict the new drug-target-disease interactions. The possible combination of this approach with network attributes, such as topological ones (e.g. different centrality measures) or the addition of more types of data, such as anatomical therapeutic chemical (ATC) codes of drugs, and continuous sequence similarities of proteins. Lastly, we are confident that performance will continue to increase as more accurate and complete input data become available.